\documentclass[conference]{IEEEtran}
\ifCLASSINFOpdf
\else
\fi

\usepackage{graphicx}
\usepackage{subcaption}
\usepackage{acronym}
\usepackage{booktabs}
\usepackage{makecell}
\usepackage{overpic}
\usepackage{float}
\usepackage[colorinlistoftodos,prependcaption,textsize=tiny]{todonotes}

\newcommand{\colordot}{%
  \tikz[baseline=-0.7ex]\node[circle,fill=green!90,minimum size=3mm] {};%
}

\begin{document}

\acrodef{CPU}{Central Processing Unit}
\acrodef{GPU}{Graphics Processing Unit}
\acrodef{NoC}{Network-on-Chip}
\acrodef{VNoC}{Vertical Network-on-Chip}
\acrodef{HNoC}{Horizontal Network-on-Chip}
\acrodef{HBM}{High-bandwidth Memory}
\acrodef{SLL}{Super Long Line}
\acrodef{AIE}{AI Engine}
\acrodef{GT}{Gigabit Transceiver}
\acrodef{FPGA}[FPGA]{Field Programmable Gate Array}
\acrodef{LUT}{Lookup Table}
\acrodef{FF}{Flip-flop}
\acrodef{MPSoC}{Multi-processor System-on-Chip}
\acrodef{QoS}{Quality-of-Service}
\acrodef{SLR}{Super Logic Region}
\acrodef{SSIT}{Stacked Silicon Interconnect Technology}
\acrodef{NIDB}{NoC Inter-die Bridge}
\acrodef{RPTR}{NoC Signal Repeater}
\acrodef{NMU}{NoC Manager Unit}
\acrodef{NSU}{NoC Subordinate Unit}
\acrodef{NPS}{NoC Packet Switch}
\acrodef{NCRB}{NoC Clock Re-convergent Buffer}
\acrodef{DRAM}{Dynamic Random-Access Memory}
\acrodef{PL}{Programmable Logics}

\title{Demystifying FPGA Hard NoC Performance}

\renewcommand{\paragraph}[1]{\noindent \textbf{#1}~}

\author{
\IEEEauthorblockN{Sihao Liu, Jake Ke, Tony Nowatzki and Jason Cong}
\IEEEauthorblockA{
University of California, Los Angeles\\
\{sihao, jakeke, tjn, cong\}@cs.ucla.edu}
}

\maketitle

\begin{abstract}
With the advent of modern multi-chiplet \ac{FPGA} architectures, vendors have begun integrating \emph{hardened} \acfp{NoC} to address the scalability, resource usage, and frequency disadvantages of soft NoCs. However, as this work shows, effectively harnessing these hardened \acp{NoC} is not trivial. It requires detailed knowledge of the microarchitecture and how it relates to the physical design of the \ac{FPGA}. Existing literature has provided in-depth analyses for \acp{NoC} in \acp{MPSoC} devices, but few studies have systematically evaluated hardened \acp{NoC} in \acp{FPGA}, which have several unique implications.

This work aims to bridge this knowledge gap by \emph{demystifying} the performance and design trade-offs of hardened \acp{NoC} on \acp{FPGA}. Our work performs detailed performance analysis of hard (and soft) \acp{NoC} under different settings, including diverse \ac{NoC} topologies, routing strategies, traffic patterns and different external memories under various NoC placements. 

In the context of Versal FPGAs, our results show that using hardened \ac{NoC} in multi-\ac{SLR} designs can reduce expensive cross-\ac{SLR} link usage by up to $30 \sim 40\%$, eliminate general-purpose logic overhead, and remove most critical paths caused by large on-chip crossbars. However, under certain aggressive traffic patterns, the frequency advantage of hardened \ac{NoC} is outweighed by the inefficiency in the network microarchitecture. We also observe suboptimal solutions from the NoC compiler and distinct performance variations between the vertical and horizontal interconnects, underscoring the need for careful design. These findings serve as practical guidelines for effectively integrating hardened \acp{NoC} and highlight important trade-offs for future FPGA-based systems.
\end{abstract}

\IEEEpeerreviewmaketitle

\section{Introduction}\label{sect:intro}

\acfp{NoC} serve as the communication backbone for connecting accelerators, memories, and I/Os across large chips~\cite{Hu2005,Park2012}. NoCs mapped to soft FPGA logic are key components in many designs~\cite{Lin2009,Udipi2012, hbm_connect, topsort}.  
However, as FPGAs become larger, soft NoCs are approaching to their limits, especially when FPGAs incorporate multiple chiplets, resulting more communication along chiplet boundaries.  
Thus, FPGA vendors, such as AMD~\cite{swarbrick2019xilinx}, Intel~\cite{esposito2023intel}, and Achronix~\cite{cairncross2023ai}, have integrated hardened NoCs.  Hard NoCs provide higher bandwidth per cross-chiplet I/O with faster clocks, and they also mitigate the resource overhead and timing closure challenges of soft NoCs~\cite{Vangal2007,Tabrizchi2019}.

However, understanding when to use the hard NoC in these systems is far from obvious. While there can be significant frequency benefits, these can be outweighed by bandwidth losses, depending on the traffic pattern and the physical layout of the design.
Making matters more complicated is the non-uniform nature of hard-NoCs.  For example, AMD's Versal NoC, as explained in Figure~\ref{fig:diag-overview}, is non-uniform along horizontal and vertical dimensions, leading to non-obvious tradeoffs for FPGA programmers.

In this work, we look at one family of FPGAs --- AMD Versal FPGAs --- and characterize the performance tradeoffs under different assumptions that represent common use cases.\footnote{We will extend our work to other families of FPGA with hardened NoCs}

To this end, we develop infrastructure to systematically evaluate a range of NoC placement, data movement patterns and \ac{QoS} scenarios on \acp{FPGA}. Our primary goal is to give insight to FPGA programmers along these lines:

\begin{figure}
    \centering
    \includegraphics[width=0.75\linewidth]{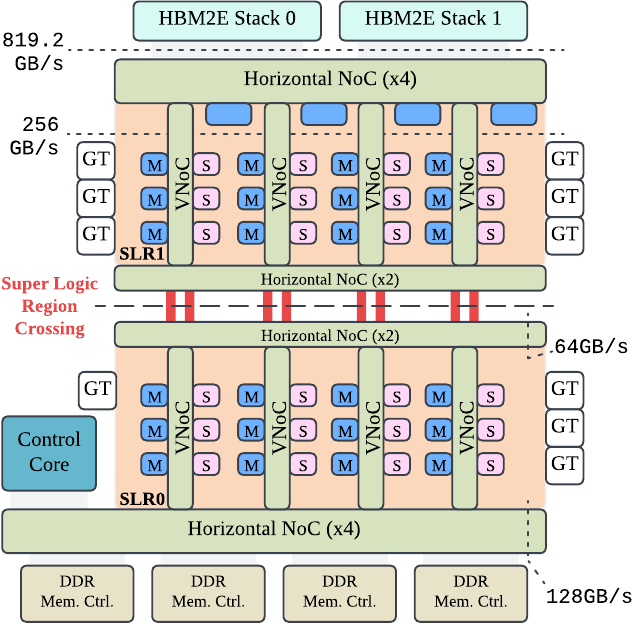}
    \caption{\small Versal FPGA Network-on-Chip Overview. 
    The NoC is non-uniform along vertical (VNoC) and horizontal (HNoC) dimensions. The VNoC bridges \ac{SLR} boundaries and reduces need for cross-die bandwidth/timing challenges.
    The \ac{HNoC} is provisioned for connecting external memory interfaces such as DRAM or \ac{HBM} controllers.}
    \vspace{-0.4cm}
    \label{fig:diag-overview}
\end{figure}


\begin{itemize}
    \item \textbf{When} and \textbf{how} to use hard \ac{NoC} efficiently? 
    \item \textbf{Which} applications and traffic profiles benefit the most from hard NoC? 
    \item \textbf{What} implementation details should users know when using a hard \ac{NoC}? 
\end{itemize}


Our main contributions are in-depth characterization of NoC's performance, soft/hard NoC comparison, profiling external memory bandwidth, and identifying NoC compiler limits.  We also synthesize guidance for programmers.
The limitations we find generate implications for future FPGA development. Finally, we release our infrastructure as an open-source NoC characterization framework for AMD FPGAs.

\paragraph{Selected Key Findings:} 
\begin{enumerate}
    \item Hard NoC read bandwidth heavily depends on the source-destination distance, while write/stream bandwidth is largely distance independent.
    \item The vertical NoC has 2 to 3$\times$ higher latency to traverse than the horizontal NoC, which may lose 20\% throughput after a single hop, while the vertical NoC can drop up to a 50\% bandwidth.
    \item The hard NoC has better scalability than the soft NoC. A 16 source/destination soft NoC (AXI-MM) fails to be mapped due to inter-die I/O, but the same design maps with no frequency loss to the hard NoC.  
    \item Soft NoCs experience frequency drops whenever they span multiple dies, even for very small networks, due to high inter-die latency.
    \item The performance degradation of hard NoC in complex traffic patterns can outweigh the relative frequency advantage over soft NoC.
    \item Achieving maximum HBM throughput is only practical in the immediate vicinity of the HBM memory controllers, and does not extend to the rest of the chip.
    \item The hard NoC compiler fails on a network size 5-times smaller than its maximum size, yet succeeds when restricting within one die.  
    This underscores the need for a more robust NoC compiler and has implications for the NoC microarchitecture.
\end{enumerate}






\section{\ac{NoC} Background}\label{sect:background}

A \ac{NoC} is a packet-based on-chip interconnect that overcomes the scalability and bandwidth limitations of traditional bus or crossbar networks. In Versal \acp{FPGA}, a \ac{NoC} is \emph{hardened} in silicon, incorporating building blocks such as \acp{NMU}, \acp{NSU}, and \acp{NPS}. These units manage packet generation/reception, protocol translation, and routing. They support features like buffering, flow control, and \acs{QoS}. Additionally, advanced stacking techniques, such as \ac{SSIT}, enable high-bandwidth connections across multiple dice, called Super Logic Regions (\acp{SLR}) in AMD \acp{FPGA}. Further architectural details can be found in the AMD official documentation \cite{xilinx_pg313} for a comprehensive description of the internals of the NoC.

\begin{figure}
    \centering
    \includegraphics[width=0.9\linewidth]{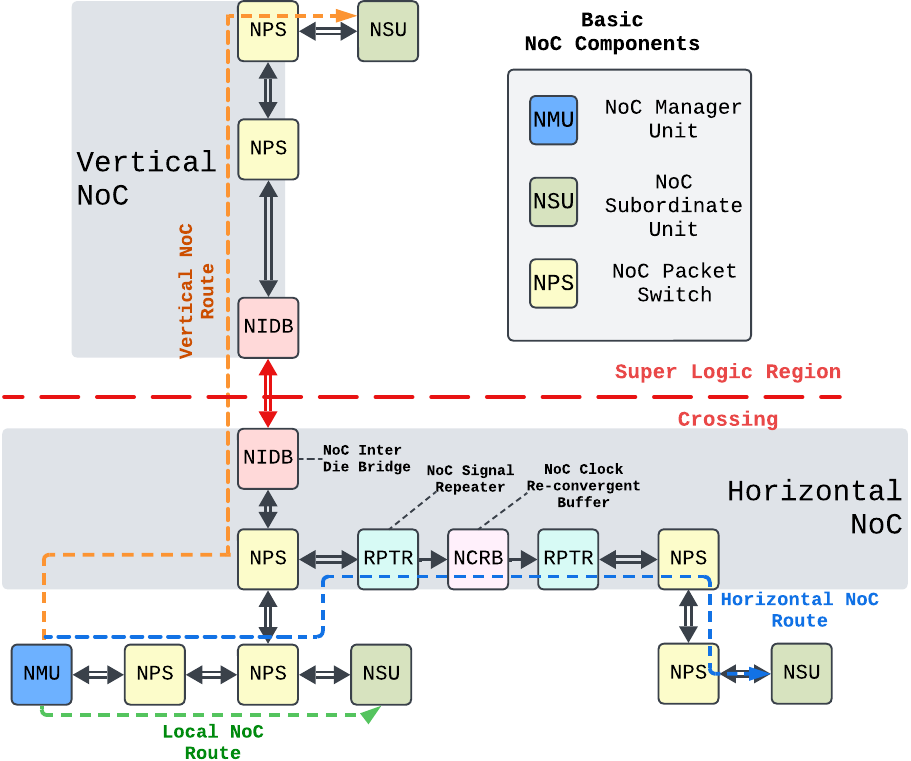}
    \caption{NoC Architecture Details and Routing Schemes}
    \vspace{-.3cm}
    \label{fig:noc_details}
\end{figure}

While these underlying components and \ac{SSIT}-based cross-SLR links serve as the fundamental infrastructure, the primary focus of this work is on the different \ac{NoC} route schemes, as depicted in Figure~\ref{fig:noc_details}. Three main routes span the device:
\begin{itemize}
\item \textbf{Local NoC Route}:
  Primarily used within a single \ac{SLR}, minimizing latency by reducing the number of switch hops. It uses nearby \acp{NPS} to directly connect local \acp{NMU} and \acp{NSU}.

\item \textbf{Vertical NoC Route}:
  Facilitates top-to-bottom data flow across \acp{SLR}, bridging dies via \ac{NIDB}. By reducing reliance on conventional cross-SLR resources, known as \acp{SLL}, the vertical route helps alleviate timing closure challenges and aligns well with vertically-oriented device resources (e.g., \ac{GT}) as in Figure~\ref{fig:diag-overview}.

\item \textbf{Horizontal NoC Route}: Supports east-west data flow using \acp{NPS}, \ac{NCRB} for clock or re-timing alignment and \ac{RPTR} for signal integrity. This route is advantageous for connecting to external memory interfaces (e.g., DDR or \ac{HBM}).
\end{itemize}

These routes collectively form the communication backbone within Versal devices, enabling robust and efficient data transport across all on-chip resources and external interfaces.

\section{Methodology} \label{sect:methodology}

\begin{figure}[t]
    \centering
    \includegraphics[width=.85\linewidth]{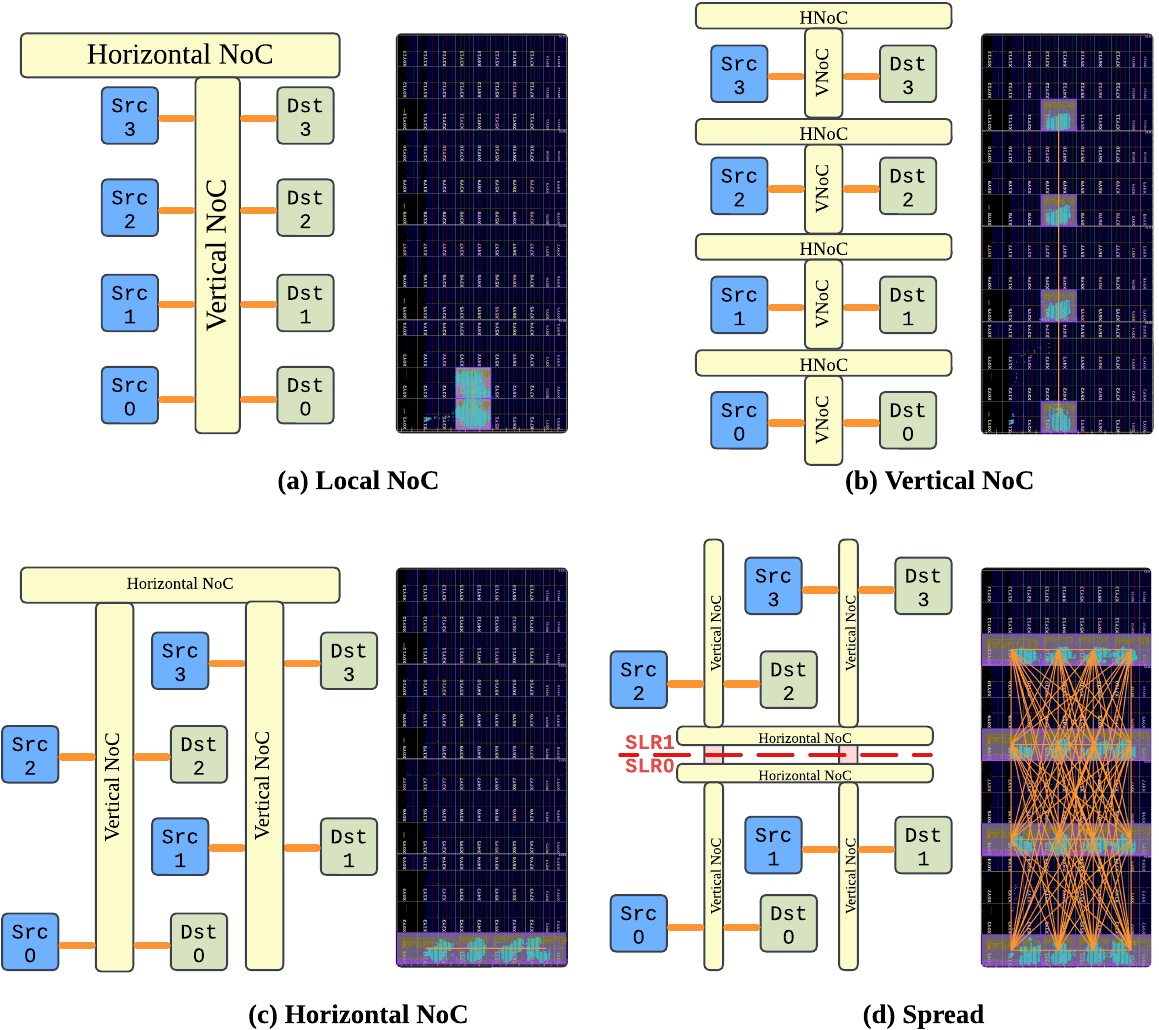}
    \caption{Different Location Setups with Floorplan}\label{fig:floorplan}
    \vspace{-.3cm}
\end{figure}

\paragraph{Benchmark Strategy \& Metrics} To characterize NoC performance across different spatial routes described in Section~\ref{sect:background}, we define four NoC placement configurations as illustrated in Figure~\ref{fig:floorplan}.

\begin{itemize}
  \item \textbf{Local}: All source and destination nodes reside in one VNoC group of one \ac{SLR} (supporting up to seven sources/destinations). This arrangement focuses on scenarios with high local density and examines the impact on throughput and latency under concentrated traffic.
  \item \textbf{HNoC}: Nodes are placed along the horizontal dimension of the NoC. This layout helps analyze the performance in terms of latency and resource consumption when communication primarily occurs over horizontal links in the same \ac{SLR}.
  \item \textbf{VNoC}: The nodes are placed along the FPGA’s vertical dimension, traversing multiple \acp{SLR} boundaries. This scenario primarily evaluates the performance implications of crossing these boundaries.
  \item \textbf{Spread}: Nodes are distributed through the chip. For instance, with four sources and destinations, a pair of each is placed on a different corner of the NoC. With sixteen pairs, they are spread out on VP1802 as shown in Figure~\ref{fig:floorplan}. This approach allows evaluation of throughput and latency under a maximally dispersed traffic scenario.
\end{itemize}

Additionally, we consider the following benchmark metrics to evaluate the system:

\begin{itemize}
  \item \textbf{Throughput (GBps)}: Measures the amount of data that the NoC can transmit within a given time frame, reflecting its overall data handling capability.
  \item \textbf{NoC Latency (ns)}: Quantifies the one-way delay from source to destination over the network, indicating how quickly the network sink can receive the requests.
  \item \textbf{FPGA Resource Utilization (\%) \& Frequency (MHz)}: Observes the efficiency of hardware resource usage.
\end{itemize}

By combining these four placement strategies with the chosen benchmark metrics, we can comprehensively assess NoC performance, hardware overhead, 
and possible operational frequency across various spatial traffic distributions.

\begin{figure}[t]
    \centering
    \includegraphics[width=.6\linewidth]{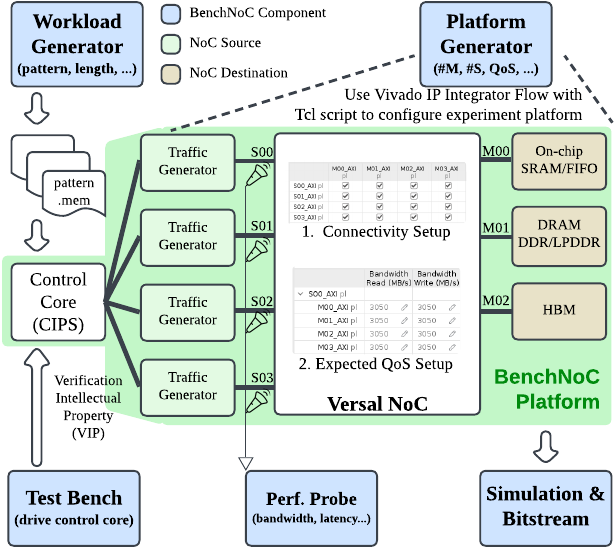}
    \caption{BenchNoC Toolchain and NoC configuration example}
    \label{fig:toolchain}
\end{figure}

\paragraph{\emph{BenchNoC} Toolchain}
The \emph{BenchNoC} toolchain automates Network-on-Chip (NoC) experiment generation and performance analysis. The toolchain comprises several key components, as illustrated in Fig.~\ref{fig:toolchain}. A \emph{Workload Generator} produces traffic patterns (e.g., nearest-neighbor, explained in Section~\ref{sect:p2p_bw}) and stores them in pattern files. These patterns are processed by a \emph{Control Core} (CIPS) that orchestrates multiple \emph{Traffic Generators}, each modeling a NoC transmitter according to the specified connectivity setup.

The \emph{Platform Generator} configures the experimental platform
by determining: (i) the number of managers/sources (\#M) and subordinates/sinks (\#S), (ii) QoS constraints, and (iii) memory configurations including on-chip SRAM/FIFOs, DDR/LPDDR DRAM, or HBM. 
It leverages Tcl scripts within the Vivado IP Integrator flow to automate both connectivity and QoS setup of AMD Vivado hard NoC IP.
During runtime, a \emph{Test Bench} drives the Control Core via Verification IP (VIP) modules, while a \emph{Performance Monitor} collects bandwidth and latency metrics.

This comprehensive framework enables systematic evaluation of NoC performance characteristics across diverse workloads and platform configurations.

\begin{table}[h]
\centering
\caption{Comparison of FPGA Resources and Specifications}
\begin{tabular}{c|c|c}
\toprule
\textbf{FPGA} & \textbf{VP1802} & \textbf{VH1782} \\
\midrule
\#SLR       & 4                 & 3 \\
\#NMU/\#NSU & 100               & 76 \\
Protocol    & AXI-Stream/AXI4-MM   & AXI4-MM \\
Memory Type & SRAM/DRAM         & HBM \\
Product  & VPK180 Board   & Alveo V80 \\
\midrule
Frequency (MHz)   & \multicolumn{2}{c}{ Design = 250MHz, \ac{NoC} = 1080MHz } \\
\midrule
Simulation Tool & \multicolumn{2}{c}{RTL Sim. by Synopsys VCS -2023.12-SP1} \\
\midrule
Physical Impl. & \multicolumn{2}{c}{AMD Vivado 2024.2}\\
\bottomrule
\end{tabular}
\label{tab:fpga_comparison}
\end{table}

\paragraph{Experiment Platform}
We employ two AMD Versal FPGAs in our experiments---the \emph{VP1802} (VPK180 board) with four SLRs, AXI-Stream/AXI4 support, and on-board SRAM/DRAM, and the \emph{VH1782} (Alveo~V80 card) with three SLRs, AXI4 support, and integrated HBM (see Table~\ref{tab:fpga_comparison}). Both devices share the \emph{-3HP} speed grade, are verified by Synopsys~VCS (V-2023.12-SP1), and are physically implemented via AMD Vivado (v2024.2). We primarily employ VH1782 for HBM-based throughput evaluations and VP1802 for all other experiments. Our block designs are developed using the Vivado IPI flow, as described in Figure~\ref{fig:toolchain}. 

\paragraph{Soft NoC Baseline}
To ensure a fair comparison with the hard NoC, we selected the functionally equivalent AMD Vivado IP \emph{AXI SmartConnect}—as our Soft NoC baseline implementation. For AXI-Stream, we similarly employed the \emph{AXI4-Stream Switch} Vivado IP. We configure these IPs with the same number of sources/destinations to mirror the data width and connectivity requirements of hard NoC IP for the purpose of apple-to-apple comparison. 

\paragraph{Workload Pattern and Protocol Characterization}
We evaluate seven traffic patterns ranging from point-to-point to one-to-all connections (Section~\ref{sect:crossbar_traffic}). Our baseline configuration uses a 512-bit data path, 4\,KB transaction size, 64\,KB total transactions per source--destination pair, and a 250\,MHz clock, except for HBM-based NMUs which operate at 400\,MHz with a 256-bit interface. We test three protocols: AXI4-MM (memory-mapped) read-only (bidirectional channels, read address request and read data response), AXI-MM write-only (write address and data request, and response), and AXI4-Stream (unidirectional data channel). QoS parameters are set to either maximize throughput for small networks (\textless 4 pairs, 16\,GB/s divided by sources) or ensure routability for larger ones (5\,MB/s).

\begin{figure}[t]
    \centering
    \begin{subfigure}[b]{0.49\linewidth}
        \centering
        \begin{overpic}[height=8cm]{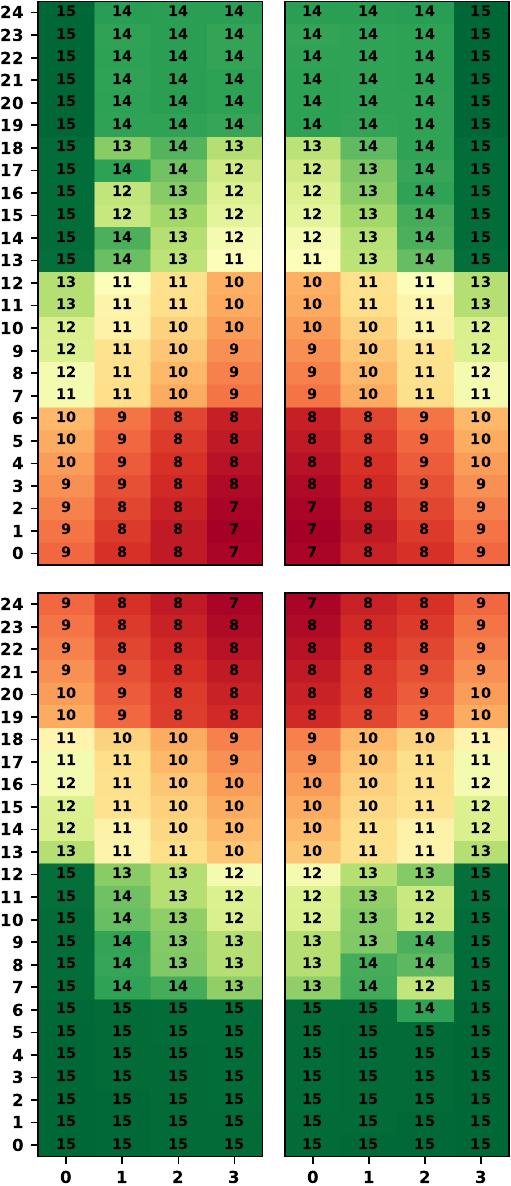}
            \put(1,-2){\tikz\node[circle,fill=green!90,minimum size=3mm]  {};}
            \put(1,100){\tikz\node[circle,fill=green!90,minimum size=3mm] {};}
            \put(41.5,-2){\tikz\node[circle,fill=green!90,minimum size=3mm]  {};}
            \put(41.5,100){\tikz\node[circle,fill=green!90,minimum size=3mm] {};}
        \end{overpic}
        \caption{Read}
        \label{fig:read_bw_heatmap}
    \end{subfigure}
    \begin{subfigure}[b]{0.49\linewidth}
        \centering
        \begin{overpic}[height=8cm]{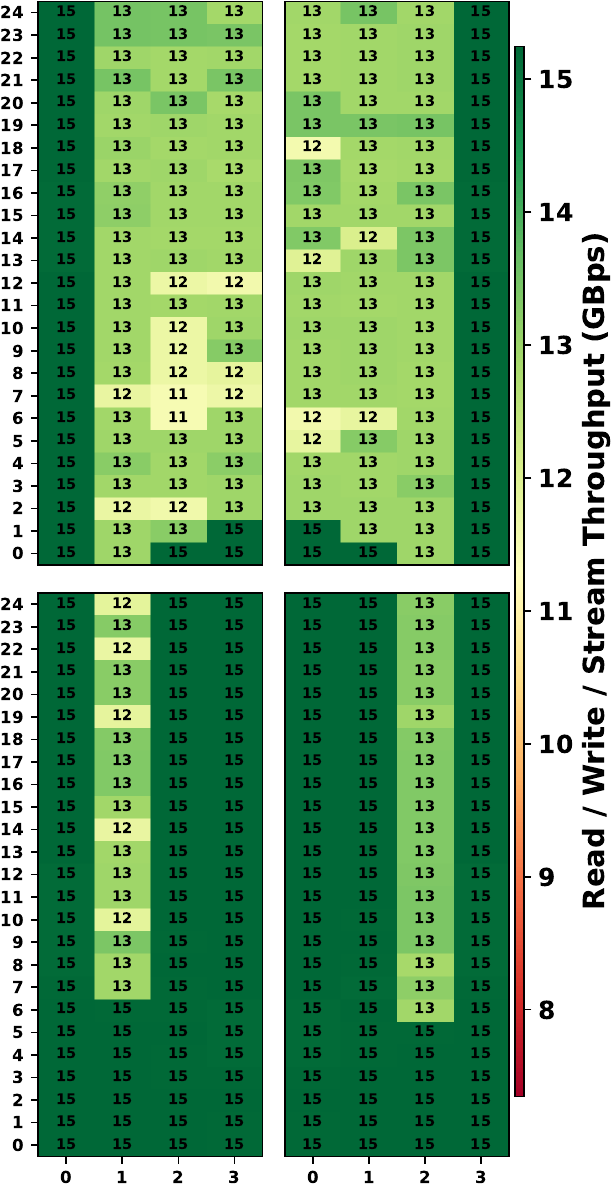}
            \put(1,-2){\tikz\node[circle,fill=green!90,minimum size=3mm]  {};}
            \put(1,100){\tikz\node[circle,fill=green!90,minimum size=3mm] {};}
            \put(41.5,-2){\tikz\node[circle,fill=green!90,minimum size=3mm]  {};}
            \put(41.5,100){\tikz\node[circle,fill=green!90,minimum size=3mm] {};}
        \end{overpic}
        \caption{Write / Stream}
        \label{fig:write_bw_heatmap}
    \end{subfigure}
    \caption{\small Heat maps of (a) AXI-MM read-only  and (b) write-only/AXI-S throughput from corner source locations (green dot) to all possible destinations on an VP1802. Warmer (red/orange) hues indicate lower throughput, while cooler (green) hues represent higher throughput.}
    \label{fig:bw_heatmaps}
    \vspace{-0.1in}
\end{figure}
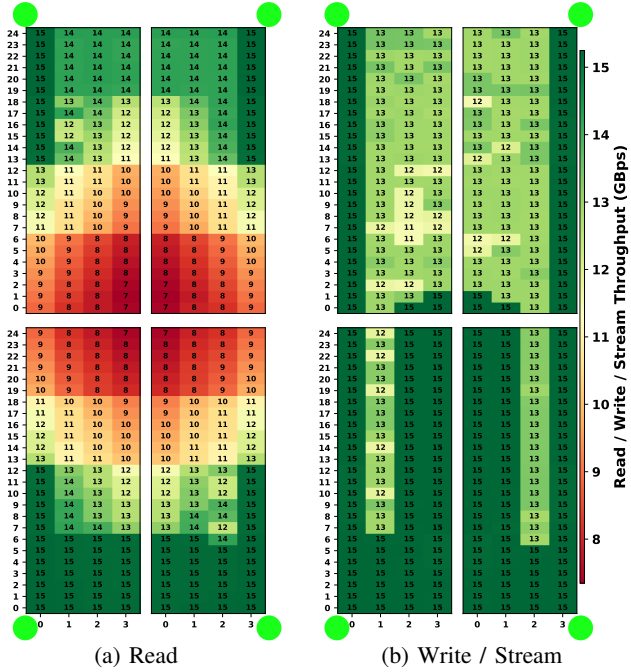

\section{Point-to-point Throughput and Latency}\label{sect:noc_bw}

In this section, we characterize how NoC throughput and latency varies with the number of hops and the placement throughout the chip, for read, write and stream traffic. In these experiments the source \ac{NMU} is 
at one of four corners (top-left, top-right, bottom-left, and bottom-right, shown in Figure \ref{fig:bw_heatmaps} as \colordot) and try every possible \ac{NSU} as a traffic destination.

\subsection{Throughput}

Figure~\ref{fig:bw_heatmaps} highlights a clear distinction of throughput heat maps between AXI-MM read-only traffic and AXI-MM write-only/AXI-S traffic, as we discuss next.

\paragraph{Read Bandwidth}
The read bandwidth is highly sensitive to communication distance, with degradation in both vertical and horizontal dimensions.
There is minor degredation per hop in both dimensions, with and more significant drops when crossing SLR boundaries, as evidenced by the three horizontal demarcations. Throughput can drop by as much as half when traveling vertically across the chip, whereas traveling horizontally sees a comparatively small reduction of about 1 to 3 GB/s. A subtle point is that in SLR0 (the SLR closest to the DDR), the read bandwidth does not exhibit a noticeable drop when traversing in the horizontal direction. This stems from the fact that SLR0 lies adjacent to the HNoC communicating with the DDR memory controller, thereby affording SLR0 additional routing resources and a larger bandwidth budget. 

\emph{Key Finding}: Significant read throughput loss with communication distance; incremental loss of 0.5 GB/s per HNoC crossing and 1-2 GB/s per VNoC crossing.

\begin{figure}
    \centering
    \begin{overpic}[height=5cm]{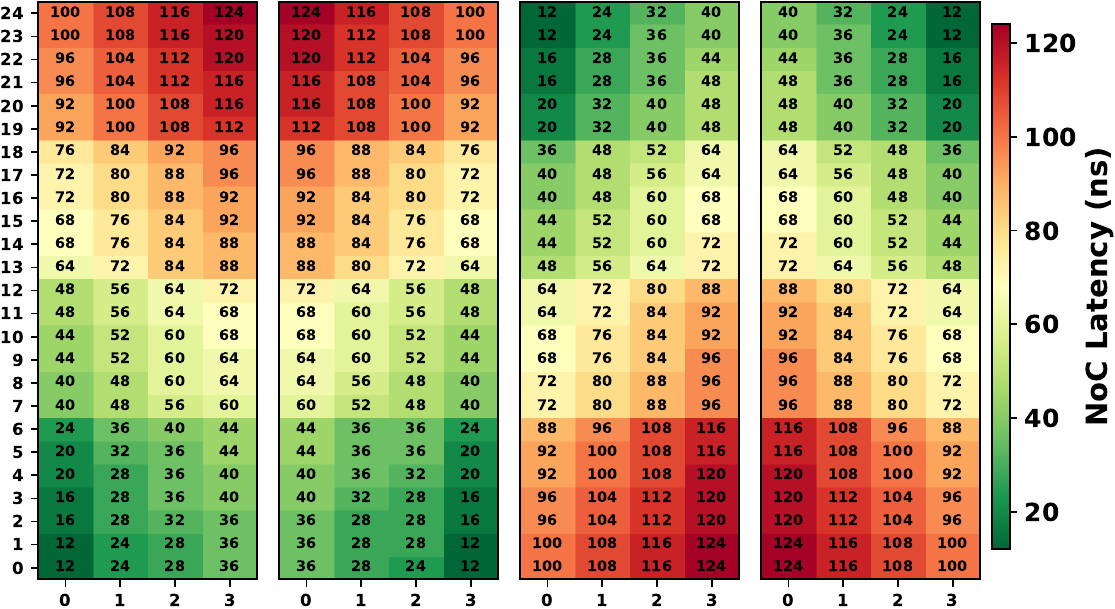}
            \put(1,-1){\tikz\node[circle,fill=green!90,minimum size=3mm]  {};}
            \put(47,55){\tikz\node[circle,fill=green!90,minimum size=3mm] {};}
            \put(43,-1){\tikz\node[circle,fill=green!90,minimum size=3mm]  {};}
            \put(84,55){\tikz\node[circle,fill=green!90,minimum size=3mm] {};}
        \end{overpic}
    \caption{\small Heat maps of NoC latency from corner source locations (green dot) to all possible destinations on VP1802. Warmer (red/orange) hues indicate higher latency, while cooler (green) hues represent lower latency.}\label{fig:latency}
    \vspace{-.5cm}
\end{figure}
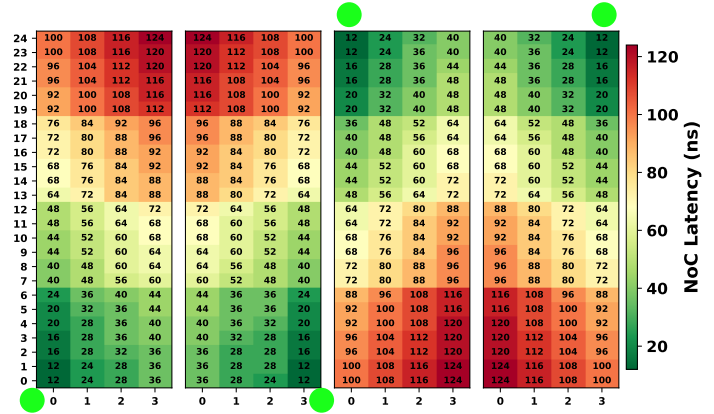

\paragraph{Write/Stream Bandwidth}
Figure~\ref{fig:write_bw_heatmap} shows the corresponding heat maps for AXI-MM write-only and AXI-S traffic, both of which avoid major throughput reductions across SLR boundaries due to effective pipelining of their simpler forwarding requests (AXI-MM write-only's response channel contains minimal traffic). Routing through the HNoC incurs a fixed bandwidth drop of approximately 3 GB/s after one hop, but no further drops occur across additional hops. This behavior is likely related to the implementation constraints of the NCRB, which only are traversed in the horizontal direction. An interesting asymmetry emerges between the top and bottom heat maps: when the starting point is at the bottom, the NoC compiler more frequently utilizes the DDR HNoC --- hence the dark green boxes in columns 2 and 3---bypassing the HNoC in PL implemented with the NCRB, which seems to suggest a deficiency in NoC Compiler (will discuss in Section~\ref{sect:compiler}). Meanwhile, the few dark green boxes visible in the top heat maps occur when that same DDR HNoC is opportunistically used for those destinations.

\emph{Key Finding}: Write and stream bandwidth is insensitive to communication distance, and there is positional asymmetry.

\subsection{Latency}

\paragraph{Hard NoC} Figure~\ref{fig:latency} shows the NoC latency (in ns) heat maps from four corner requesters (marked as \colordot) of VP1802 to all possible destination locations. Unlike the asymmetric throughput results, these latency patterns are nearly identical for all source locations: as the number of NoC hops increases, latency rises accordingly. The largest increases appear vertically every two jumps, especially when crossing SLR boundaries. Horizontal traversal adds a more gradual increment in latency.

\paragraph{Soft NoC} In contrast, the Soft NoC exhibits more uniform latencies, with consistent delays between all source-destination pairs at approximately 10 cycles. Comparing with hard NoC, we observe that intra-SLR communication in hard NoC achieves similar latency (around 10 cycles at 250MHz). However, for cross-SLR designs, Soft NoC faces additional frequency challenges (discussed in \ref{sect:noc_util_freq}).

\emph{Key Finding}: For single-SLR, latency-tolerant designs, Soft NoC matches Hard NoC latency, but consumes additional FPGA resources.
\section{Effect of Traffic Patterns on Hard vs Soft NoC} \label{sect:crossbar_traffic}

\begin{figure}[t]
    \centering
    \includegraphics[width=0.8\linewidth]{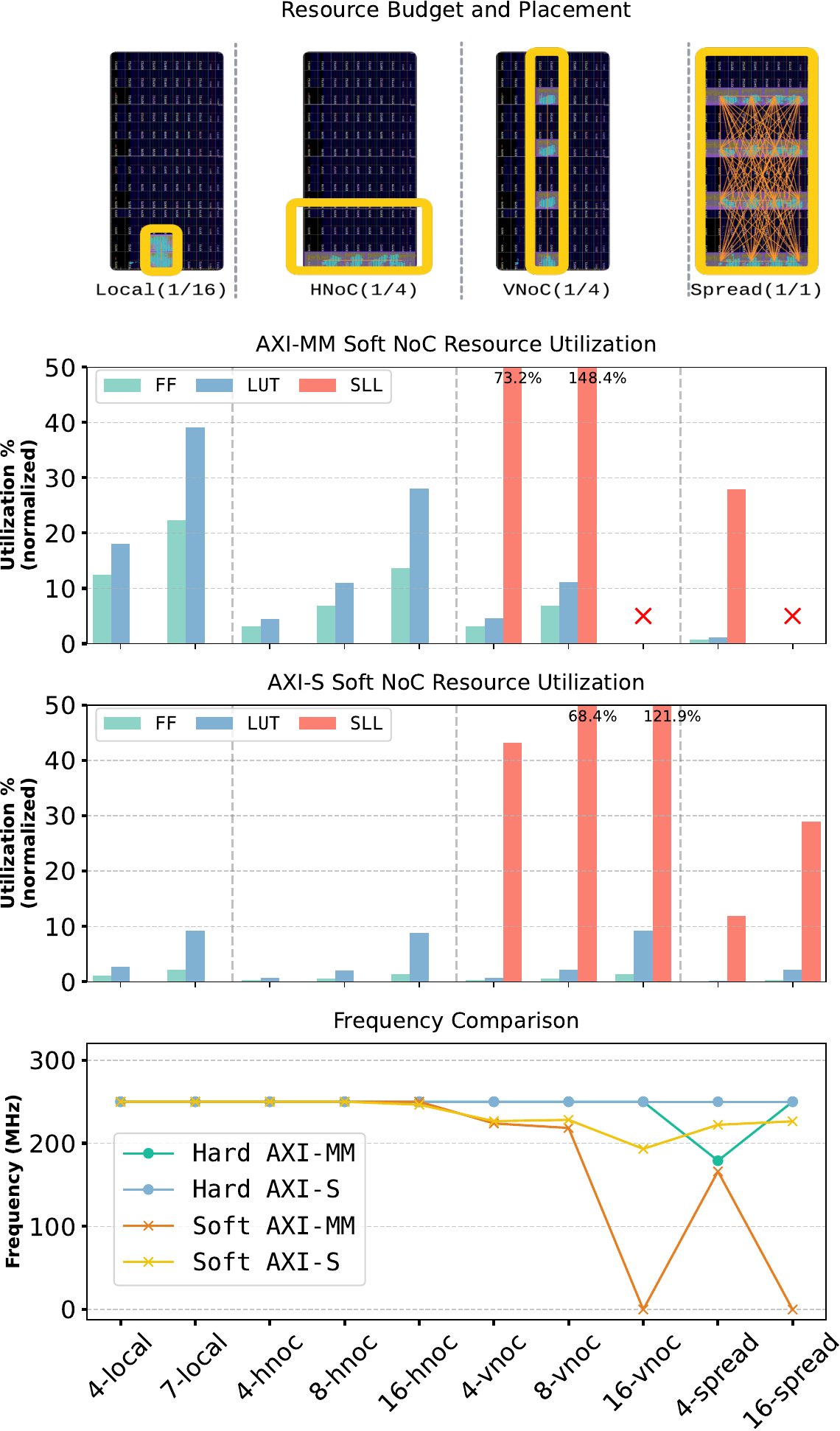}
    \caption{\small Soft NoC resource utilization and SLR-crossing (upper), and frequency comparison of soft and hard NoC (lower) for both AXI-MM and AXI-S, with various crossbar size (4, 7, 8, 16) and (local, HNoC, VNoC, spread). Large soft NoCs consume substantial resources and extra SLR-crossing wires, some cases failing to route.}
    \vspace{-0.5cm}
    \label{fig:util_freq}
\end{figure}

Crossbars are commonly used in accelerator designs, either when all-to-all communication is needed,
or when there is a change in patterns over time. Indeed, a key motivation for designers to adopt hard NoC is to replace soft crossbars (referred to here as “soft NoCs”) in their designs. 


In this section, we present a comprehensive analysis of resource utilization, frequency, and throughput between a soft NoC (SmartConnect IP, as in Section~\ref{sect:methodology}) and hard NoC, for different traffic patterns. Section~\ref{sect:noc_util_freq} first shows the substantial advantage hard NoCs have over soft NoCs in resource utilization and frequency. Section~\ref{sect:p2p_bw} and Section~\ref{sect:access_all_bw} then detail how throughput changes under varying levels and distributions of traffic congestion under different traffic patterns. We begin by analyzing the point-to-point pattern in a crossbar-based scenario, since it constitutes a fundamental phase in more complex algorithms. We then expand our evaluation to a scenario involving access to all destinations. Finally, we compare against an “Ideal Crossbar,” which is the soft NoC if it could achieve the targeted frequency (250MHz).

\subsection{Resource Utilization and Frequency} \label{sect:noc_util_freq}

We compare frequency and resource usage of the ten crossbar configurations with a soft NoC, normalizing resource usage by each design’s fraction of the FPGA (e.g., 1/16 for Local, 1/4 for VNoC and HNoC, as shown at top of Figure~\ref{fig:util_freq}). 

In Figure~\ref{fig:util_freq}, the second and third plot shows normalized resource utilization (\ac{LUT}, \ac{FF}, and \ac{SLR}-crossing wires) for \emph{soft} \ac{NoC} implementations of AXI-MM and AXI4-stream. We omit \emph{hard} \ac{NoC} since it does not consume those resources. The bottom plot compares achieved frequency under the 250 MHz target. Along the x-axis, the data points progress from Local, HNoC, VNoC to Spread placements, each tested at increasing crossbar sizes. 

For Local and HNoC comparison, placement is constrained to one SLR, so the soft NoC has no extra SLR-crossing wires. The LUT and FF consumption indicates that when all logic is packed into a small Local region, utilization can become concentrated and suboptimal, expanding placement to the entire SLR better distributes resources and reduce congestion. Both soft NoC and hard NoC meet the target frequency. For VNoC and Spread, placement involves multiple SLRs. Therefore, SLR-crossing utilization increase significantly with size for soft NoC. For medium size of soft NoC placed as VNoC, the normalized SLL resource reaches 148\% means that only two VNoC crossbars can be implemented using soft NoC, while hard NoC support up to 4. For a crossbar size of 16, SLR-crossing wire demands exceed 100\%, leading to failed routing.

In contrast, the hard NoC maintains high frequencies across all sizes and placements, underscoring the advantages of offloading NoC functionality to dedicated silicon, which preserves timing and eliminates the LUT/FF overhead.

\emph{Key Finding}: Hard NoC outperforms the soft NoC in both resource utilization and frequency, especially for the multi-SLR crossing. However, it can be degraded under certain specialized traffic patterns (explained in Section~\ref{sect:p2p_bw}).

\begin{figure}[t]
    \centering
    \includegraphics[width=\linewidth]{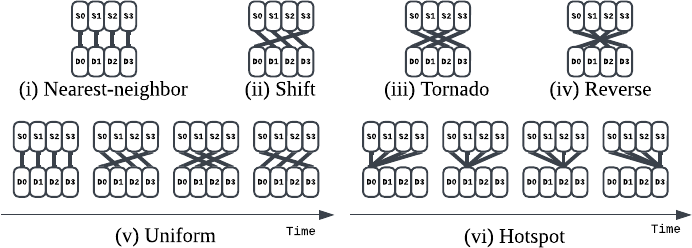}
    \caption{\small	 Traffic patterns: (i) nearest-neighbor: each source sends to the nearest destination, (ii) shift: each source sends to the next indexed destination; (iii) tornado: sources send halfway around the network; (iv) reverse: sources send to destinations mirrored across the network center; (v) uniform: sources send to all destinations in sequence; (vi) hotspot: all sources transmit to every destination at once.}
    \vspace{-0.5cm}
    \label{diag:pattern}
\end{figure}

\begin{figure*}[t]
    \centering
    \begin{subfigure}[b]{0.24\textwidth}
        \centering
        \includegraphics[width=\linewidth]{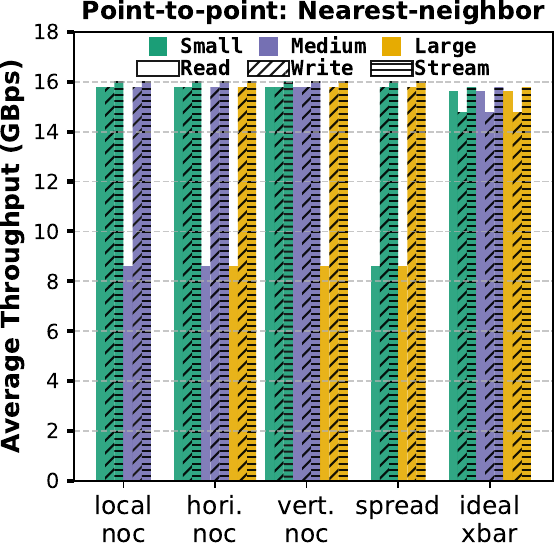}
        \caption{Nearest-neighbor}
        \label{fig:local_bw}
    \end{subfigure}
    \hfill
    \begin{subfigure}[b]{0.24\textwidth}
        \centering
        \includegraphics[width=\linewidth]{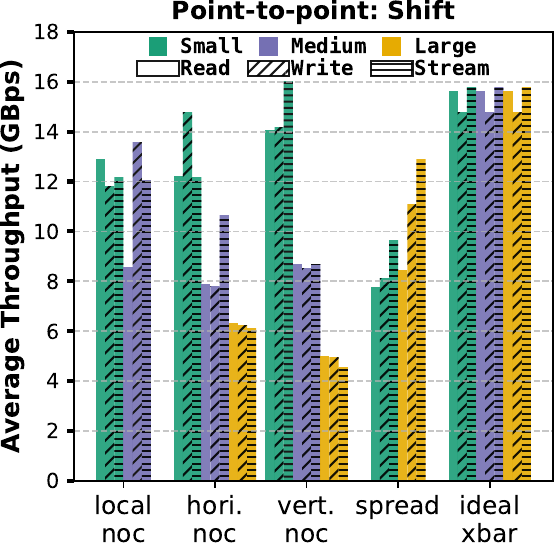}
        \caption{Shift}
        \label{fig:shift_bw}
    \end{subfigure}
    \hfill
    \begin{subfigure}[b]{0.24\textwidth}
        \centering
        \includegraphics[width=\linewidth]{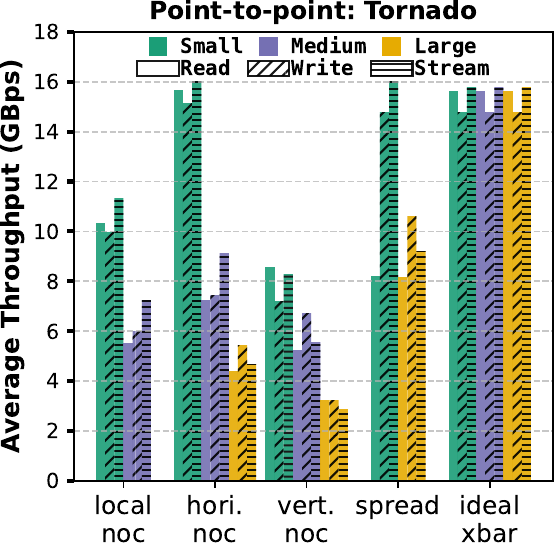}
        \caption{Tornado}
        \label{fig:tornado_bw}
    \end{subfigure}
    \hfill
    \begin{subfigure}[b]{0.24\textwidth}
        \centering
        \includegraphics[width=\linewidth]{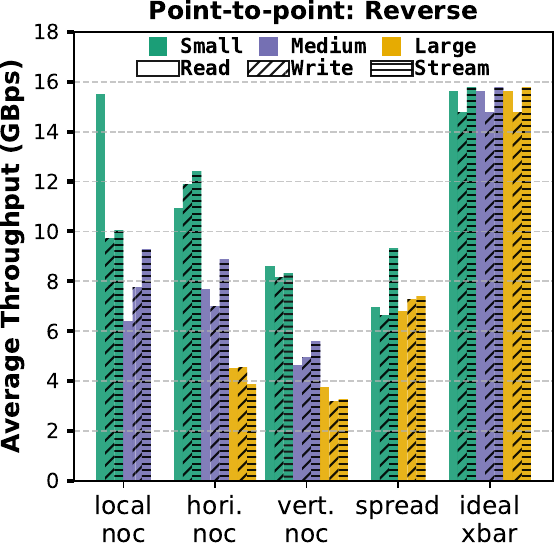}
        \caption{Reverse}
        \label{fig:reverse_bw}
    \end{subfigure}
    \caption{Average source throughput under point-to-point traffic patterns across multiple crossbar sizes and NoC placements.}
    \vspace{-0.2cm}
    \label{fig:p2p_patterns}
\end{figure*}

\begin{figure*}[t]
    \centering
    \begin{subfigure}[b]{0.32\textwidth}
        \centering
        \includegraphics[width=.8\linewidth]{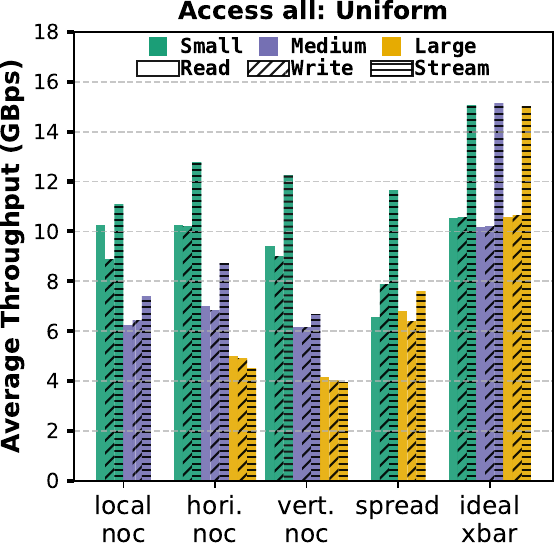}
        \caption{Uniform}
        \label{fig:uniform_bw}
    \end{subfigure}
    \hfill
    \begin{subfigure}[b]{0.32\textwidth}
        \centering
        \includegraphics[width=.8\linewidth]{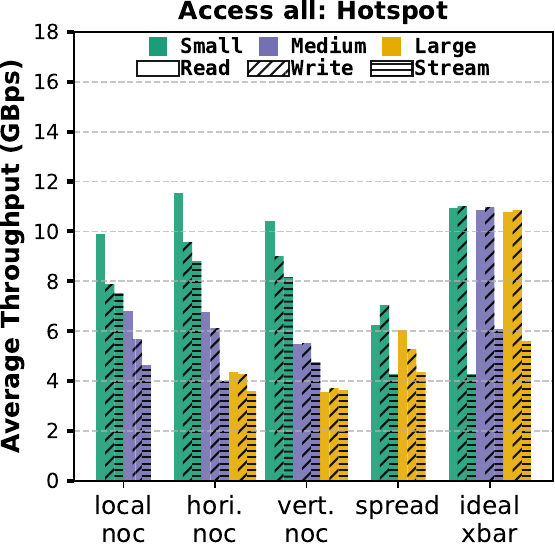}
        \caption{Hotspot}
        \label{fig:hotspot_bw}
    \end{subfigure}
    \hfill
    \begin{subfigure}[b]{0.32\textwidth}
        \centering
        \includegraphics[width=.8\linewidth]{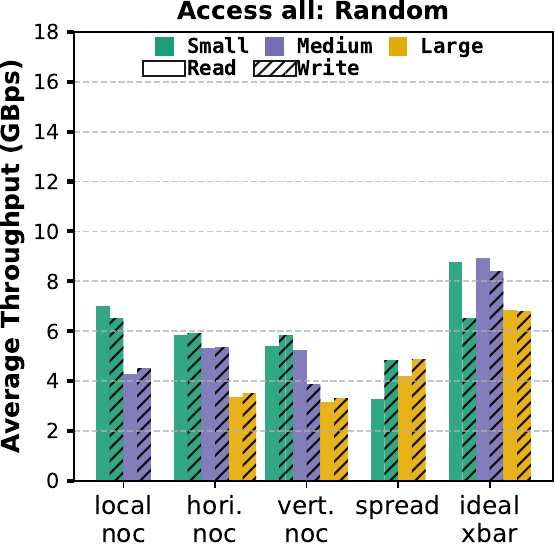}
        \caption{Random}
        \label{fig:random_bw}
    \end{subfigure}
    \caption{Average source throughput under access-all traffic patterns across multiple crossbar sizes and NoC placements.}
    \vspace{-.6cm}
    \label{fig:all_patterns_2}
\end{figure*}

\subsection{Throughput of Point-to-point Traffic Patterns}\label{sect:p2p_bw}

\paragraph{Nearest-Neighbor}
Figure~\ref{diag:pattern}(i) illustrates the nearest-neighbor traffic pattern, where each source sends data to its physically closest destination, resulting in zero congestion. Figure~\ref{fig:local_bw} shows that most configurations achieve the expected maximum NoC link throughput of about 16\,GB/s. However, AXI-MM read traffic experiences an unexpected drop to roughly half this value, likely due to suboptimal static configurations or NoC architecture. Because this issue may affect other traffic patterns as well, we do not focus on read-only throughput in subsequent analysis.

\paragraph{Shift}
Figure~\ref{diag:pattern}(ii) shows the shift traffic pattern, where each source sends data to the next destination index in a “ring.” Unlike nearest-neighbor traffic, shift requires multiple hops, causing some connections to share links in crossbar configurations. As indicated by Figure~\ref{fig:shift_bw}, Local, HNoC, and VNoC placements yield lower throughput than the ideal crossbar, with more significant drops as crossbar size increases. By contrast, spread placement can perform differently depending on routing constraints: in small-spread scenarios, certain central links remain idle, whereas large-spread setups utilize those links more effectively.

\paragraph{Tornado}
Figure~\ref{diag:pattern}(iii) shows the tornado pattern where sources send data halfway around the network. HNoC placement drives horizontal communication between network sides, while VNoC emphasizes vertical traffic across SLR boundaries, with spread placement distributing this load. Though tornado typically yields lower throughput than shift patterns due to increased congestion (Figure~\ref{fig:tornado_bw}), small-HNoC and small-spread configurations maintain near-ideal performance through optimized static routing.

\paragraph{Reverse}
Figure~\ref{diag:pattern}(iv) shows the reverse traffic pattern, where each source sends data to the destination mirrored across the center of the network, thus placing heavy pressure on central NoC links. Except for occasional zero-congestion corner cases resulting from static routing, the average throughput resembles the tornado pattern (see Figure~\ref{fig:reverse_bw}). In the worst-case congestion scenarios, the effective bandwidth drops to approximately 40\% of the small crossbar size, with this significant performance degradation primarily attributed to bottlenecks in the central links.

\emph{Key Findings}: While patterns like nearest-neighbor can achieve near-ideal bandwidth, challenging patterns like Tornado and Reverse can be affected by link-sharing congestion, especially for vertical NoC. Large-spread placement or using horizontal NoC may mitigate such performance loss.

\subsection{Access-all Traffic Patterns}\label{sect:access_all_bw}

\begin{figure*}[t]
    \centering
    \includegraphics[width=.85\textwidth]{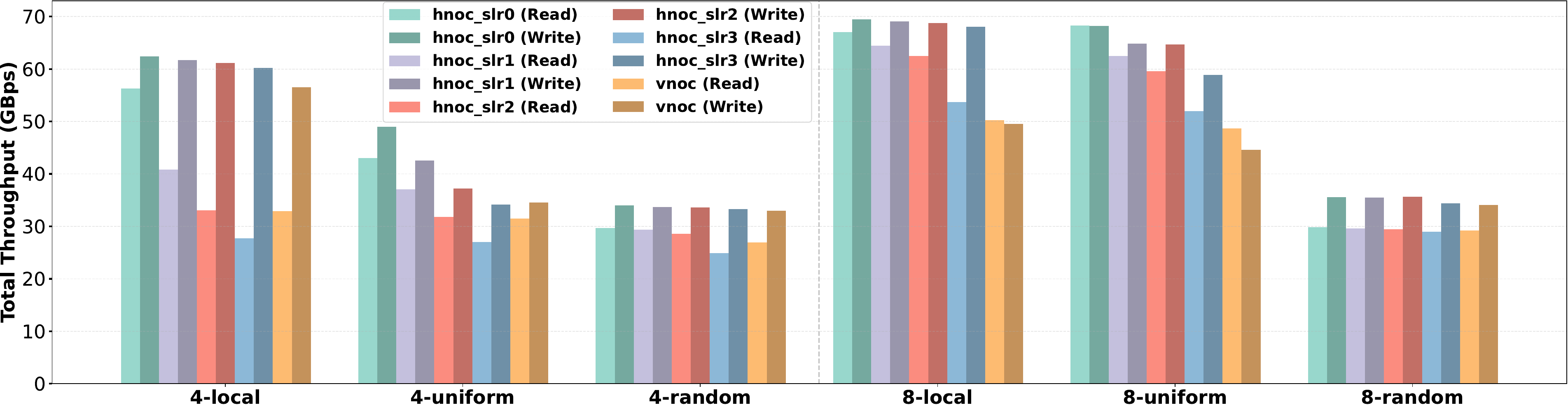}
    \caption{\small DRAM throughput: 4/8 sources, VNoC/HNoC placement, AXI-MM R/W patterns}
    \vspace{-.5cm}
    \label{fig:dram_bw}
\end{figure*}

\paragraph{Uniform}
Figure~\ref{diag:pattern}(v) illustrates the uniform traffic pattern, where each source progressively sends data to all destinations, transitioning from a local-like to a shift-like distribution. As shown in Figure~\ref{fig:uniform_bw}, throughput stays near ideal levels under various NoC placements, suggesting effective load balancing even at higher congestion states.

\paragraph{Hotspot}
Figure~\ref{diag:pattern}(vi) depicts hotspot traffic, where every source transmits to the same destination simultaneously before moving on. This synchronization raises congestion, lowering average source throughput relative to uniform (Figure~\ref{fig:hotspot_bw}). However, each destination individually observes high throughput, and the gap between Versal NoC and ideal crossbar remains modest.

\paragraph{Random}
Under random traffic, each source selects a new destination for every 4\,KB AXI-MM transaction. We exclude AXI-S for practicality. This pattern can exacerbate congestion, driving the network toward a more pronounced bottleneck (Figure~\ref{fig:random_bw}).

\emph{Key Findings}: Balanced uniform and random traffic remains near ideal for read, with up to $\sim 20$\% write/stream penalty; Hotspot traffic saturates destination at 40\% penalty, reducing source bandwidth.
\section{External Memory Accesses}

The Versal NoC is essential for Programmable Logics (PL) to DRAM/HBM communication. We demonstrate NoC's ability to achieve peak memory throughput and analyze performance across configurations to establish design guidelines.

\subsection{Dynamic Random-Access Memory (DRAM)}

We examine three traffic patterns (nearest-neighbor, uniform, and random) introduced in Section~\ref{sect:crossbar_traffic}, each representative of common DRAM access scenarios. Unlike ``nearest-neighbor'' in Section~\ref{sect:p2p_bw}, here it means each source communicates with its nearest memory controller via a strided access pattern. We conduct experiments using four or eight sources (limited by eight bidirectional NoC links to DRAM) and consider five NoC placements: four horizontal variants (one per SLR, noting that SLR~0 is closest to DDR and SLR~3 is the farthest) plus a VNoC configuration. These experiments aim to measure DRAM bandwidth utilization and analyze how source-to-DRAM distance impacts overall performance.

Figure~\ref{fig:dram_bw} shows that the best-performing setup reaches nearly 70\,GB/s, matching practical DRAM bandwidth expectations. Additionally, write performance often surpasses read throughput. Consistent with Section~\ref{sect:noc_bw}, sources placed closer to DRAM (e.g., SLR~0) achieve higher throughput, whereas distant placements suffer. Increasing the number of sources from four to eight further enhances parallelism in the DRAM controllers, though using more than eight sources eventually saturates the NoC links.

\emph{Key Finding}: Placing sources closer to the DRAM controllers sustains higher throughput by up to 50\% more bandwidth from SLR crossing loss, while additional sources improve parallelism but eventually saturate the NoC links.

\subsection{High-bandwidth Memory (HBM)}

\begin{figure}[t]
   \centering
   \includegraphics[width=0.75\linewidth]{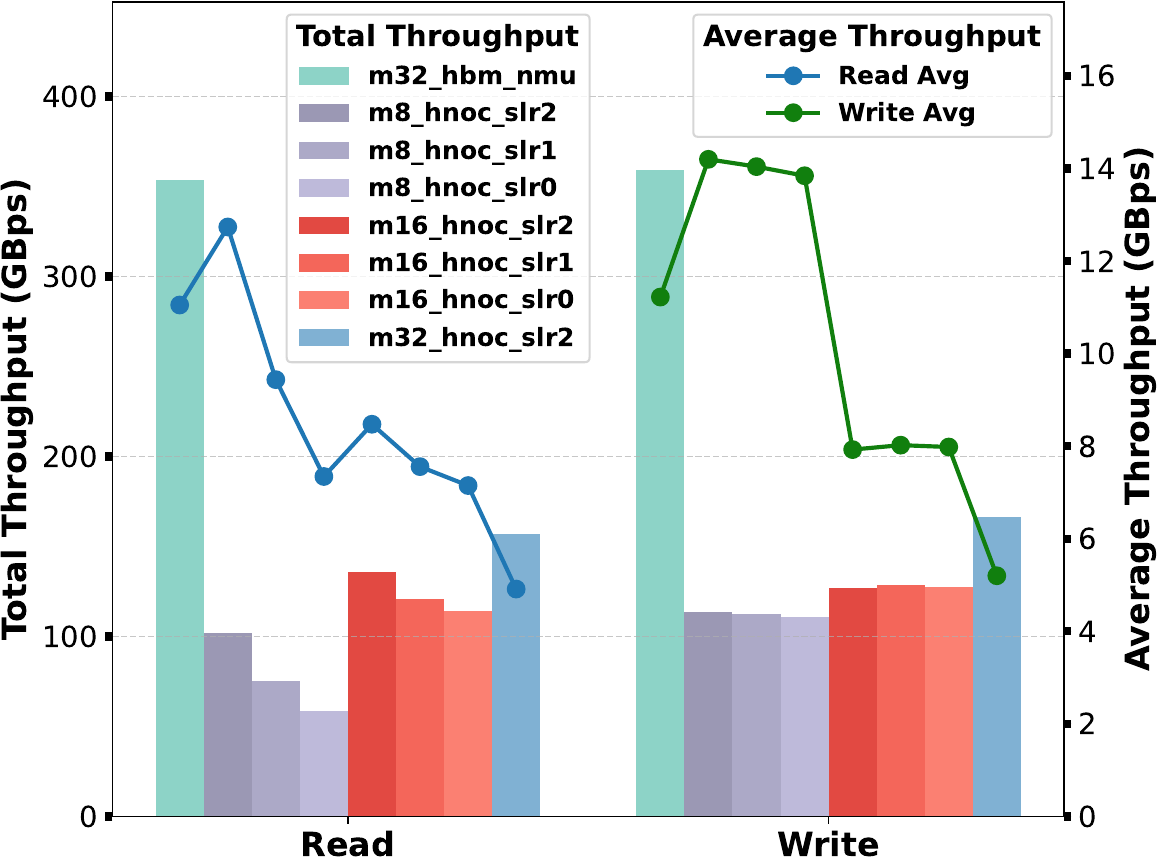}
   \caption{\small HBM throughput (bars: total GB/s, lines: average GB/s) for R/W neighbor traffic using PL NMUs vs HBM NMUs (8/16/32 sources). HBM NMUs (32, one-stack) reach 354\,GB/s vs theoretical 409.6\,GB/s; PL NMUs achieve \textless 50\% throughput.}
   \vspace{-.5cm}
   \label{fig:hbm_bw}
\end{figure}

We evaluate HBM performance using nearest-neighbor traffic patterns with AXI-MM read-only and write-only operations. Our analysis encompasses three network configurations (8, 16, and 32 sources) and two distinct placement strategies:

\begin{itemize}
   \item \textbf{HBM NMU Configuration}: Sources are placed by using 256-bit HBM NMUs in nearest SLR to one HBM stack
   \item \textbf{Horizontal NoC (HNoC) Configuration}: Sources are distributed along the horizontal NoC dimension across SLR 0 - 2, with SLR 2 maintaining closest physical proximity to HBM stacks. For the 32-source HNoC configuration in SLR 2, we utilize 24 NMUs from SLR 2 supplemented by 8 NMUs from SLR 1, because each non-bottom SLR has up to 24 NMUs/NSUs.
\end{itemize}

Figure~\ref{fig:hbm_bw} compares total source throughput (bars, left y-axis) and average source throughput (lines, right y-axis) for 32 HBM NMUs on a single HBM stack versus NMUs distributed in the Programmable Logic (PL) region. The HBM NMUs achieve 354\,GB/s for both read and write, approaching 86\% of the 409.6\,GB/s theoretical limit for a single stack. By contrast, PL NMUs reach about 160\,GB/s even when optimally placed, and suffer further drops as they move farther from the HBM controllers, illustrating how bandwidth is severely constrained by increased distance and NoC constraints.

\emph{Key Finding}: 
Full HBM bandwidth can only be delivered to the nearest SLR, which limits the design scale to a fraction of that which can be mapped practically while achieving full bandwidth.  

\section{\ac{NoC} Compiler}\label{sect:compiler}

\begin{figure}[t]
    \centering
    \includegraphics[width=.7\linewidth]{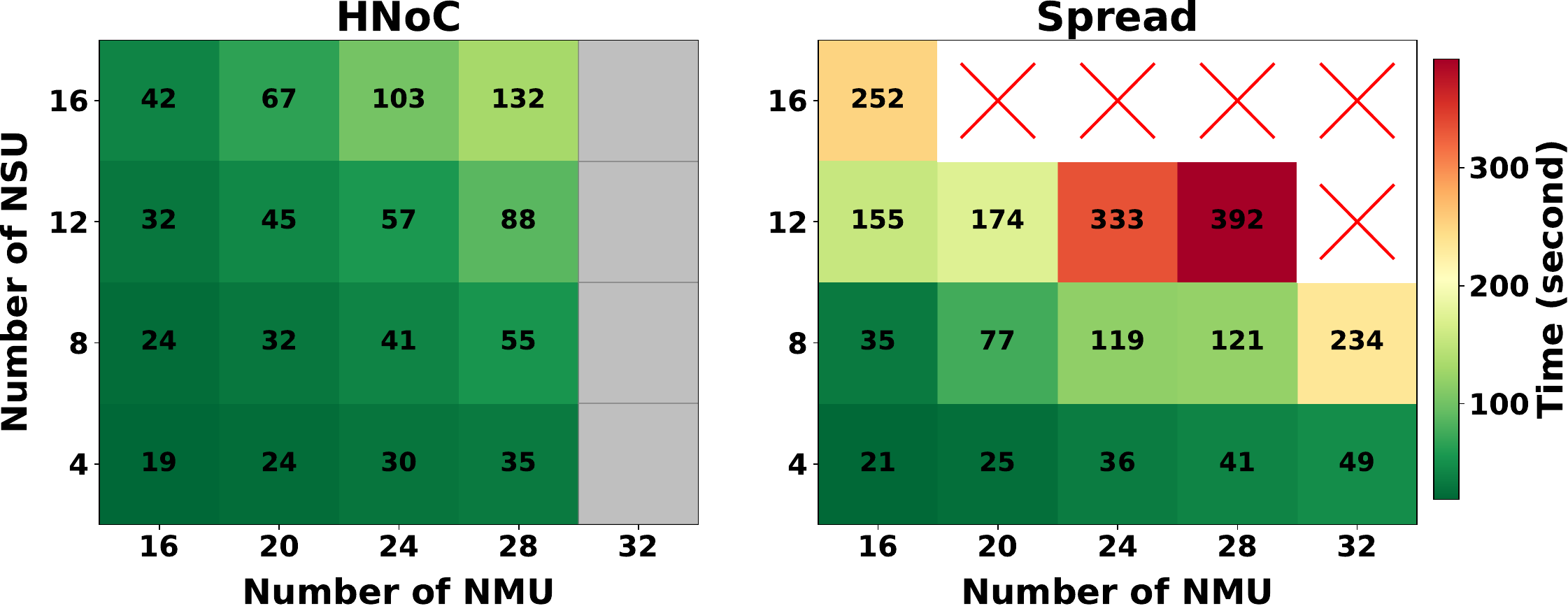}
    \caption{NoC compiler time on HNoC and Spread placement (\textcolor{gray!50}{\rule{0.5em}{0.5em}}: not enough \#NMU; \textcolor{red}{\(\times\)}: timeout at 5\,MB/s link bandwidth)}\label{fig:compiler}
    \vspace{-.5cm}
\end{figure}

The NoC compiler~\cite{versal_noc} is a proprietary tool 
that leverages a SAT-based engine (e.g., {\em minisat}\cite{sorensson2010minisat}) to find routing 
solutions that meet desired bandwidth, latency, and traffic requirements. 
In our experiments, we focus on scenarios where the requested bandwidth 
is deliberately set to a very low level (5MB/s) such that the NoC compiler's 
ability to find a route is more reflective of microarchitectural constraints than bandwidth constraints.  We vary the size of the network in number of source/destination pairs to observe the compilation time and success rate. Figure~\ref{fig:compiler} shows the results, with compile times of up to a few minutes for larger designs.

Surprisingly, when using the whole FPGA (spread), several designs fail, that succeeded when using a single die (HNoC), e.g. 20 NMUs, 16 NSUs.  For the spread design, the largest network (28 NMUs, 12 NSUs) uses only about $1/4 \sim 1/5$ of the maximum size for the FPGA (100 NMUs, 100 NSUs). This indicates that there are microarchitecture constraints that either make mapping these sizes impossible, or are too problematic for the current SAT-based technique.  This imposes a strict design constraint for FPGA programmers, especially for cross-chip networks.

\emph{Key Finding}: While there is potential to reach larger scale networks, microarchitecture and compiler limitations should be addressed.
\section{Related Work} 

\paragraph{Compare to ASIC/CPU NoCs} These employ fixed architectures and achieve high efficiency through optimized routing under strict area and power constraints~\cite{Dally2001,Bjerregaard2006,Kumar2002,Kim2008}. In contrast, FPGA NoCs encounter lower clock frequencies, higher resource overhead from reconfigurable fabrics~\cite{Lin2009,Kapre2006,Abdelfattah2012}, and limited adaptability in partially hardened designs like AMD Versal~\cite{xilinx_pg313}. Nevertheless, FPGA NoCs offer flexibility for custom communication, while hardened NoCs enhance performance but lack application-specific or FPGA optimization~\cite{Tabrizchi2019, Papamichael2012, nagalaxmi2024fpga, Anuradha2024, jerger2022chip}.

\paragraph{Benchmarking Methodologies} Performance evaluation commonly relies on synthetic traffic patterns~\cite{grecu2007towards,Murali2004,Hu2003,Pham2012, gratz2006implementation, versal_noc_latency_sim}, but FPGA NoCs also require scrutiny of resource utilization and hardware constraints~\cite{Udipi2012, Vantrease2011, nguyen2023spades}. Although hybrid approaches have been proposed, many remain insufficient for FPGA-specific challenges~\cite{Bjerregaard2005}.

\paragraph{Unresolved Challenges in FPGA NoCs} Adapting ASIC-oriented routing algorithms and developing adaptive strategies for dynamic workloads remain difficult~\cite{Guerrier2000,Tabrizchi2019, Papamichael2012, liu2022overgen}. The lack of standardized benchmarks and efficient resource management adds complexity~\cite{gupta2020run, Kapre2006}. Balancing soft NoC flexibility with hardened components is still an open question~\cite{Lin2009, Abdelfattah2012, ramesh2021fpga}. Recent works~\cite{guo2021autobridge, alonso2021elastic, agarwal2024analysis} optimize multi-die FPGA designs effectively with automatic interconnect, they do not leverage NoC architectures.
\section{Conclusion}

In this work, we have investigated the advantages, limitations, 
and design considerations of the hardened NoC in Versal devices. We conclude 
by answering the three key questions posed in Section~\ref{sect:intro}:

\noindent\textbf{Q1: When and how to use hard \ac{NoC} efficiently?}
\begin{itemize}
    \item \textbf{Frequency and SLR-crossing savings:} 
    By offloading expensive \ac{SLL} required for \ac{SLR} crossings, the Versal NoC significantly alleviates timing closure issues. This benefit is especially pronounced in large, multi-\ac{SLR} designs, where hardened NoC can save an additional \textbf{30--40\%} of crossing resources compared to soft crossbars.
    \item \textbf{Weigh against negative factors:} The congestion and frequency benefits must be weighed against bandwidth losses for long-distance reads and pathological communication patterns.
    \item \textbf{When used for external memory:} 
    Near-peak external memory bandwidth can be \textbf{twice} higher in the nearest \ac{SLR} compared to the farthest one, so careful design consideration should be made so that bandwidth can be consumed near memory controllers.  HBM bandwidth must be consumed near HBM controllers.
\end{itemize}

\noindent\textbf{Q2: Which traffic profiles benefit the most?}
\begin{itemize}
    \item \textbf{Bandwidth-heavy, write-dominated, or streaming workloads:} AXI-S and AXI-MM writes are less sensitive to \ac{SLR} boundaries, making high-throughput streaming data or data-logging good candidates for the hard NoC.
    \item \textbf{Nearest neighbor or ring-based dataflow:} As shown in Section~\ref{sect:crossbar_traffic}, these patterns benefit most from the hardened NoC. Highly parallel designs also mitigate multi-\ac{SLR} overhead via localized dataflow.
\end{itemize}

\noindent\textbf{Q3: What implementation details should users know?}
\begin{itemize}
    \item \textbf{Vertical vs.\ horizontal routing:} AXI-MM reads can degrade by up to \textbf{50\%} across multiple \ac{SLR} boundaries, so plan data paths carefully.
    \item \textbf{Traffic contention} The hard NoC sees \textbf{60\%} throughput drops (e.g., ``Reverse'' pattern) under heavy contention.
    \item \textbf{Compiler’s handling of crossbars and routing:} For large, distributed crossbars with minimum bandwidth needs, the NoC compiler may fail to find a solution or produce a suboptimal one. Consider manual constraints or segmenting crossbars when dealing with large networks.
\end{itemize}

Overall, the hardened \ac{NoC} simplifies high-performance designs by reducing timing 
and resource overhead. Nonetheless, traffic patterns, NMU placement and inter-\ac{SLR} routing remain pivotal considerations for achieving peak throughput.

\bibliographystyle{IEEEtran}

\begin{thebibliography}{10}
\providecommand{\url}[1]{#1}
\csname url@samestyle\endcsname
\providecommand{\newblock}{\relax}
\providecommand{\bibinfo}[2]{#2}
\providecommand{\BIBentrySTDinterwordspacing}{\spaceskip=0pt\relax}
\providecommand{\BIBentryALTinterwordstretchfactor}{4}
\providecommand{\BIBentryALTinterwordspacing}{\spaceskip=\fontdimen2\font plus
\BIBentryALTinterwordstretchfactor\fontdimen3\font minus \fontdimen4\font\relax}
\providecommand{\BIBforeignlanguage}[2]{{%
\expandafter\ifx\csname l@#1\endcsname\relax
\typeout{** WARNING: IEEEtran.bst: No hyphenation pattern has been}%
\typeout{** loaded for the language `#1'. Using the pattern for}%
\typeout{** the default language instead.}%
\else
\language=\csname l@#1\endcsname
\fi
#2}}
\providecommand{\BIBdecl}{\relax}
\BIBdecl

\bibitem{Hu2005}
J.~Hu and R.~Marculescu, ``Energy- and performance-aware mapping for regular noc architectures,'' \emph{IEEE Transactions on Computer-Aided Design of Integrated Circuits and Systems}, vol.~24, no.~4, pp. 551--562, 2005.

\bibitem{Park2012}
D.~Park, S.~Ramanathan, T.~Theocharides, V.~Narayanan, and M.~Irwin, ``Reliability and energy savings in on-chip networks using distributed spare routers,'' in \emph{Proceedings of the Design, Automation \& Test in Europe Conference \& Exhibition (DATE)}, 2012, pp. 1341--1344.

\bibitem{Lin2009}
C.~Lin, L.~Ni, and J.~Wu, ``Fault-tolerant routing in 3d networks-on-chip,'' \emph{IEEE Transactions on Computers}, vol.~60, no.~6, pp. 824--836, 2009.

\bibitem{Udipi2012}
A.~Udipi, N.~Muralimanohar, R.~Balasubramonian, N.~P. Jouppi, and S.~W. Keckler, ``Toward scalable, energy-efficient, bus-based on-chip networks,'' in \emph{Proceedings of the IEEE 18th International Symposium on High Performance Computer Architecture (HPCA)}, 2012, pp. 1--12.

\bibitem{hbm_connect}
\BIBentryALTinterwordspacing
Y.-k. Choi, Y.~Chi, W.~Qiao, N.~Samardzic, and J.~Cong, ``Hbm connect: High-performance hls interconnect for fpga hbm,'' in \emph{The 2021 ACM/SIGDA International Symposium on Field-Programmable Gate Arrays}, ser. FPGA '21.\hskip 1em plus 0.5em minus 0.4em\relax New York, NY, USA: Association for Computing Machinery, 2021, p. 116–126. [Online]. Available: \url{https://doi.org/10.1145/3431920.3439301}
\BIBentrySTDinterwordspacing

\bibitem{topsort}
W.~Qiao, L.~Guo, Z.~Fang, M.-C.~F. Chang, and J.~Cong, ``Topsort: A high-performance two-phase sorting accelerator optimized on hbm-based fpgas,'' in \emph{2022 IEEE 30th Annual International Symposium on Field-Programmable Custom Computing Machines (FCCM)}, 2022, pp. 1--1.

\bibitem{swarbrick2019xilinx}
\BIBentryALTinterwordspacing
I.~Swarbrick, D.~Gaitonde, S.~Ahmad, B.~Gaide, and Y.~Arbel, ``Network-on-chip programmable platform in versal\textsuperscript{\texttrademark} acap architecture,'' in \emph{Proceedings of the 2019 ACM/SIGDA International Symposium on Field-Programmable Gate Arrays}, ser. FPGA '19.\hskip 1em plus 0.5em minus 0.4em\relax New York, NY, USA: Association for Computing Machinery, 2019, p. 212–221. [Online]. Available: \url{https://doi.org/10.1145/3289602.3293908}
\BIBentrySTDinterwordspacing

\bibitem{esposito2023intel}
B.~Esposito, ``Intel agilex{\textregistered} 9 direct rf-series fpgas with integrated 64 gsps data converters,'' in \emph{2023 IEEE Hot Chips 35 Symposium (HCS)}.\hskip 1em plus 0.5em minus 0.4em\relax IEEE Computer Society, 2023, pp. 1--35.

\bibitem{cairncross2023ai}
A.~Cairncross, B.~Henry, C.~Chalmers, D.~Reid, J.~Shipton, J.~Fowler, L.~Corrigan, and M.~Ashby, ``Ai benchmarking on achronix speedster{\textregistered} 7t fpgas,'' \emph{White Paper}, 2023.

\bibitem{Vangal2007}
S.~Vangal, J.~Howard, G.~Ruhl, S.~Dighe, H.~Wilson, J.~Tschanz, D.~Finan, A.~Singh, T.~Jacob, S.~Robinson \emph{et~al.}, ``An 80-tile 1.28tflops network-on-chip in 65nm {CMOS},'' in \emph{IEEE International Solid-State Circuits Conference Digest of Technical Papers (ISSCC)}, 2007, pp. 98--99.

\bibitem{Tabrizchi2019}
M.~Tabrizchi, A.~Afzali-Kusha, and M.~Pedram, ``Revisiting noc performance modeling by exploiting advanced flow control,'' in \emph{Proceedings of the Design, Automation \& Test in Europe Conference \& Exhibition (DATE)}, 2019, pp. 1523--1528.

\bibitem{xilinx_pg313}
\BIBentryALTinterwordspacing
A.~Inc., \emph{Versal Adaptive SoC Programmable Network on Chip and Integrated Memory Controller}, Nov. 2024, product Guide PG313. [Online]. Available: \url{https://docs.amd.com/r/en-US/pg313-network-on-chip}
\BIBentrySTDinterwordspacing

\bibitem{versal_noc}
I.~Swarbrick, D.~Gaitonde, S.~Ahmad, B.~Jayadev, J.~Cuppett, A.~Morshed, B.~Gaide, and Y.~Arbel, ``Versal network-on-chip (noc),'' in \emph{2019 IEEE Symposium on High-Performance Interconnects (HOTI)}, 2019, pp. 13--17.

\bibitem{sorensson2010minisat}
N.~S{\"o}rensson, ``Minisat 2.2 and minisat++ 1.1,'' \emph{A short description in SAT Race}, vol. 2010, 2010.

\bibitem{Dally2001}
\BIBentryALTinterwordspacing
W.~J. Dally and B.~Towles, ``Route packets, not wires: On-chip interconnection networks,'' in \emph{Proceedings of the 38th Annual Design Automation Conference (DAC)}, 2001, pp. 684--689. [Online]. Available: \url{https://ieeexplore.ieee.org/document/935594}
\BIBentrySTDinterwordspacing

\bibitem{Bjerregaard2006}
T.~Bjerregaard and S.~Mahadevan, ``A survey of research and practices of network-on-chip,'' \emph{ACM Computing Surveys}, vol.~38, no.~1, pp. 1--51, 2006.

\bibitem{Kumar2002}
S.~Kumar, A.~Jantsch, M.~Millberg, J.~Öberg, J.~Soininen, M.~Forsell, K.~Tiensyrjä, and A.~Hemani, ``A network on chip architecture and design methodology,'' in \emph{Proceedings of the IEEE Computer Society Annual Symposium on VLSI}, 2002, pp. 105--112.

\bibitem{Kim2008}
J.~Kim, J.~Balfour, and W.~J. Dally, ``Flattened butterfly topology for on-chip networks,'' \emph{IEEE Computer Architecture Letters}, 2008.

\bibitem{Kapre2006}
N.~G. Kapre, ``Packet-switched on-chip fpga overlay networks,'' Ph.D. dissertation, California Institute of Technology, 2006.

\bibitem{Abdelfattah2012}
M.~S. Abdelfattah and V.~Betz, ``Take the highway: Design for router-to-router networks on fpgas,'' in \emph{International Conference on Field Programmable Logic and Applications}, 2012.

\bibitem{Papamichael2012}
M.~K. Papamichael and J.~C. Hoe, ``Connect: Re-examining conventional wisdom for designing nocs in the context of fpgas,'' in \emph{International Symposium on Field Programmable Gate Arrays}, 2012.

\bibitem{nagalaxmi2024fpga}
T.~Nagalaxmi, E.~S. Rao, and P.~ChandraSekhar, ``Fpga-based implementation and verification of hybrid security algorithm for noc architecture,'' \emph{Analog Integrated Circuits and Signal Processing}, vol. 121, no.~1, pp. 13--23, 2024.

\bibitem{Anuradha2024}
P.~Anuradha, P.~Majumder, K.~Sivaraman, N.~A. Vignesh, S.~A. Jayakar, A.~Selvaraj, S.~Mallik, A.~Al-Rasheed, M.~Abbas, and B.~O. Soufiene, ``Enhancing high-speed data communications: Optimization of route controlling network on chip implementation,'' \emph{IEEE Access}, vol.~12, pp. 123\,514--123\,528, 2024.

\bibitem{jerger2022chip}
N.~E. Jerger, T.~Krishna, and L.-S. Peh, \emph{On-chip networks}.\hskip 1em plus 0.5em minus 0.4em\relax Springer Nature, 2022.

\bibitem{grecu2007towards}
C.~Grecu, A.~Ivanov, P.~Pande, A.~Jantsch, E.~Salminen, U.~Ogras, and R.~Marculescu, ``Towards open network-on-chip benchmarks,'' in \emph{First International Symposium on Networks-on-Chip (NOCS'07)}.\hskip 1em plus 0.5em minus 0.4em\relax IEEE, 2007, pp. 205--205.

\bibitem{Murali2004}
S.~Murali and G.~D. Micheli, ``Sunmap: A tool for automatic topology selection and generation for nocs,'' in \emph{Proceedings of the 41st Annual Design Automation Conference (DAC)}, 2004, pp. 914--919.

\bibitem{Hu2003}
J.~Hu and R.~Marculescu, ``Energy-aware mapping for tile-based noc architectures under performance constraints,'' in \emph{Proceedings of the Asia and South Pacific Design Automation Conference (ASP-DAC)}, 2003, pp. 233--239.

\bibitem{Pham2012}
P.-H. Pham, P.~Maidee, and J.~Peng, ``Design space exploration for fpga-based network-on-chip implementations,'' in \emph{International Symposium on Field-Programmable Custom Computing Machines}, 2012.

\bibitem{gratz2006implementation}
P.~Gratz, C.~Kim, R.~McDonald, S.~W. Keckler, and D.~Burger, ``Implementation and evaluation of on-chip network architectures,'' in \emph{2006 International Conference on Computer Design}.\hskip 1em plus 0.5em minus 0.4em\relax IEEE, 2006, pp. 477--484.

\bibitem{versal_noc_latency_sim}
\BIBentryALTinterwordspacing
I.~Lang, N.~Kapre, and R.~Pellizzoni, ``Worst-case latency analysis for the versal noc network packet switch,'' in \emph{Proceedings of the 15th IEEE/ACM International Symposium on Networks-on-Chip}, ser. NOCS '21.\hskip 1em plus 0.5em minus 0.4em\relax New York, NY, USA: Association for Computing Machinery, 2021, p. 55–60. [Online]. Available: \url{https://doi.org/10.1145/3479876.3481593}
\BIBentrySTDinterwordspacing

\bibitem{Vantrease2011}
D.~Vantrease, R.~Schreiber, M.~Monchiero \emph{et~al.}, ``Corona: System implications of emerging nanophotonic technology,'' \emph{ACM SIGARCH Computer Architecture News}, 2011.

\bibitem{nguyen2023spades}
T.~Nguyen, Z.~Blair, S.~Neuendorffer, and J.~Wawrzynek, ``Spades: A productive design flow for versal programmable logic,'' in \emph{2023 33rd International Conference on Field-Programmable Logic and Applications (FPL)}.\hskip 1em plus 0.5em minus 0.4em\relax IEEE, 2023, pp. 65--71.

\bibitem{Bjerregaard2005}
T.~Bjerregaard and J.~Sparsø, ``Implementation of guaranteed services in the mango clockless network-on-chip,'' \emph{IEE Proceedings - Computers and Digital Techniques}, vol. 152, no.~2, pp. 217--227, 2005.

\bibitem{Guerrier2000}
P.~Guerrier and A.~Greiner, ``A generic architecture for on-chip packet-switched interconnections,'' in \emph{Proceedings of the Design, Automation and Test in Europe Conference and Exhibition (DATE)}, 2000, pp. 250--256.

\bibitem{liu2022overgen}
S.~Liu, J.~Weng, D.~Kupsh, A.~Sohrabizadeh, Z.~Wang, L.~Guo, J.~Liu, M.~Zhulin, R.~Mani, L.~Zhang \emph{et~al.}, ``Overgen: Improving fpga usability through domain-specific overlay generation,'' in \emph{2022 55th IEEE/ACM International Symposium on Microarchitecture (MICRO)}.\hskip 1em plus 0.5em minus 0.4em\relax IEEE, 2022, pp. 35--56.

\bibitem{gupta2020run}
A.~Gupta, S.~Ahmed, A.~K. Jain, Y.~Arbel, A.~Morshed, and D.~Schultz, ``Run-time reconfiguration of noc in xilinx acap architecture,'' in \emph{2020 13th International Workshop on Network on Chip Architectures (NoCArc)}.\hskip 1em plus 0.5em minus 0.4em\relax IEEE, 2020, pp. 1--6.

\bibitem{ramesh2021fpga}
G.~Ramesh and S.~Prabhu, ``Fpga implementation of 3d noc using anti-hebbian for multicast routing algorithm,'' in \emph{Micro-Electronics and Telecommunication Engineering: Proceedings of 4th ICMETE 2020}.\hskip 1em plus 0.5em minus 0.4em\relax Springer, 2021, pp. 301--313.

\bibitem{guo2021autobridge}
\BIBentryALTinterwordspacing
L.~Guo, Y.~Chi, J.~Wang, J.~Lau, W.~Qiao, E.~Ustun, Z.~Zhang, and J.~Cong, ``Autobridge: Coupling coarse-grained floorplanning and pipelining for high-frequency hls design on multi-die fpgas,'' in \emph{The 2021 ACM/SIGDA International Symposium on Field-Programmable Gate Arrays}, ser. FPGA '21.\hskip 1em plus 0.5em minus 0.4em\relax New York, NY, USA: Association for Computing Machinery, 2021, p. 81–92. [Online]. Available: \url{https://doi.org/10.1145/3431920.3439289}
\BIBentrySTDinterwordspacing

\bibitem{alonso2021elastic}
T.~Alonso, L.~Petrica, M.~Ruiz, J.~Petri-Koenig, Y.~Umuroglu, I.~Stamelos, E.~Koromilas, M.~Blott, and K.~Vissers, ``Elastic-df: Scaling performance of dnn inference in fpga clouds through automatic partitioning,'' \emph{ACM Transactions on Reconfigurable Technology and Systems (TRETS)}, vol.~15, no.~2, pp. 1--34, 2021.

\bibitem{agarwal2024analysis}
P.~Agarwal, T.~Kumar~Garg, and A.~Kumar, ``Analysis of 3d noc router chip on different fpga for minimum hardware and fast switching,'' \emph{National Academy Science Letters}, vol.~47, no.~1, pp. 35--39, 2024.

\end{thebibliography}

\end{document}